# Screened Coulombic Orientational Correlations in Dilute Aqueous Electrolytes

*Luc Belloni[1]\*, Daniel Borgis[2,3] and Maximilien Levesque[3]*

[1]LIONS, NIMBE, CEA, CNRS, Université Paris-Saclay, 91191 Gif-sur-Yvette, France

[2]Maison de la Simulation, USR 3441 CNRS-CEA-Université Paris-Saclay, 91191 Gif-sur-Yvette, France

[3]PASTEUR, Département de chimie, École normale supérieure, PSL University, Sorbonne Université, CNRS, 75005 Paris, France

\*luc.belloni@cea.fr

Abstract: The ion-induced long-range orientational order between water molecules recently observed in second harmonic scattering experiments and illustrated with large scale molecular dynamics simulations is quantitatively explained using the Ornstein-Zernike integral equation approach of liquid physics. This general effect, not specific to hydrogen-bonding solvents, is controlled by electroneutrality condition, dipolar interactions and dielectric+ionic screening. As expected, all numerical theories recover the well-known analytical expressions established 40 years ago.

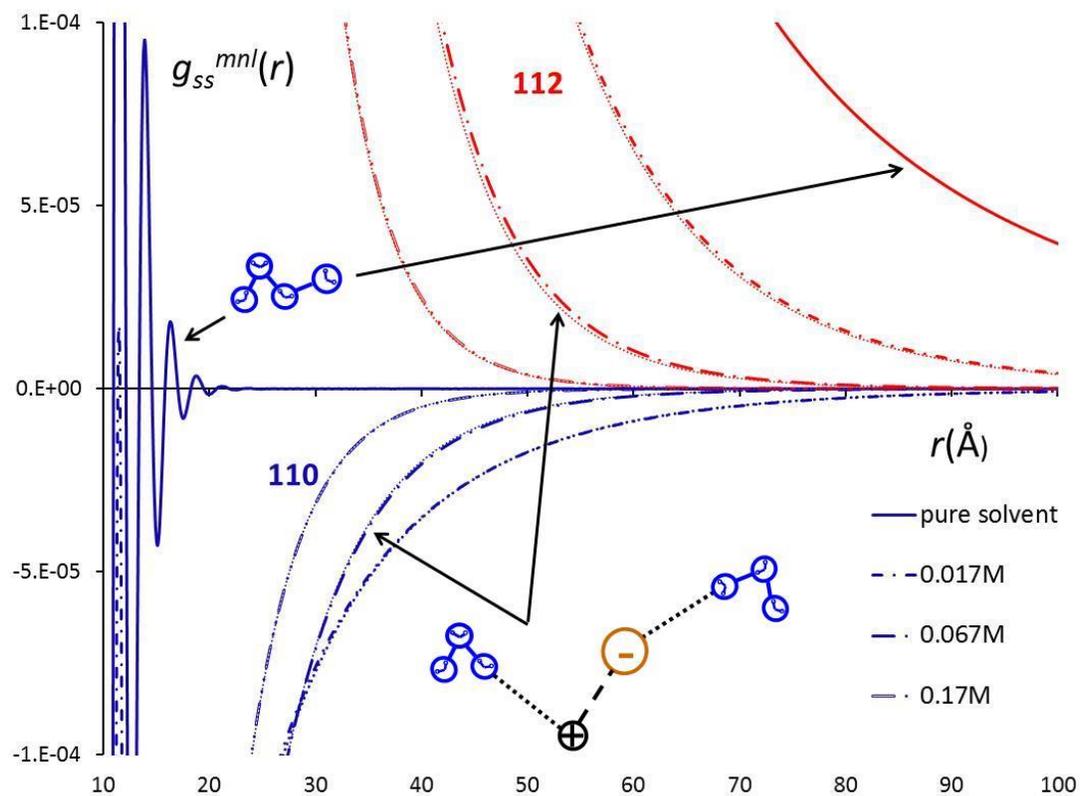

The existence of long range orientational correlations among water molecules induced by the addition of dilute salts has been the subject of recently renewed interest. Experimentally, femtosecond elastic second harmonic scattering exhibits an increase in particular angular correlations, compared to pure water, for salinity range starting from 10μM [1]. This generic effect, reproduced for different salt compositions and ionic valences, depends only on the ionic strength. Numerically, molecular dynamics simulations were performed for NaCl aqueous solutions in the 10mM concentration range using simulation cells of size 10-20nm containing hundreds of thousand water molecules [1,2]. They revealed the existence of very low but long-range correlations in the simulation box.

The theoretical study of correlations in ionic solutions with an explicit treatment of the molecular solvent on an equal statistical mechanical footing with the ions (as opposed to the primitive model picture of continuous dielectric solvent) has a long history in the literature since the late 1970's. Analytical limiting laws valid at low salinity and large separation [3,4] as well as systematic numerical resolutions of the liquid physics Ornstein-Zernike (OZ) approach with standard Mean Spherical Approximation or HyperNetted Chain (HNC) integral equation approximations have covered various types of molecular solvent, from the pure dipolar case [5,6,7] up to the more realistic site-site representation [8,9,10,11,12]. In the present context, it is fruitful to recall right away the limiting law for the solvent-solvent (*ss*) orientational correlations established by Hoye and Stell in 1978 [3,4] that seems to have been forgotten in the recent literature:

$$g_{ss}(12) \equiv g_{ss}(\vec{r}_{12}, \Omega_{12}) \underset{\substack{\text{large } r \\ \text{low salinity}}}{\approx} \frac{(\varepsilon-1)^2 \kappa_D^2}{12\pi\rho_s y \varepsilon} \frac{e^{-\kappa_D r}}{r} \Delta(12) + \frac{(\varepsilon-1)^2}{4\pi\rho_s y \varepsilon} \frac{1+\kappa_D r + \tfrac{1}{3}(\kappa_D r)^2}{r^3} e^{-\kappa_D r} D(12)$$

(1.1)

where $\rho_s$ is the number density of the solvent, $\varepsilon$ its dielectric constant, $y = \frac{4\pi}{9}\rho_s \mu_s^2 L_B$ the standard dimensionless parameter which characterizes the strength of the dipolar coupling ($\mu_s e$ is the solvent dipole, $L_B = \frac{e^2}{4\pi\varepsilon_0 kT}$ the Bjerrum length at temperature *T in vacuum*), and $\kappa_D = \sqrt{8\pi L_B I / \varepsilon}$ the usual Debye screening constant in that electrolyte of ionic strength *I*. The angular dependence of the pair distribution function in (1.1) involves the first rotational invariants or spherical harmonics $\Delta(12) = \hat{\mu}_1.\hat{\mu}_2$ and $D(12) = 3(\hat{r}_{12}.\hat{\mu}_1)(\hat{r}_{12}.\hat{\mu}_2) - \hat{\mu}_1.\hat{\mu}_2$ (the hat symbol defines unitary vectors). The coefficient in front of $\Delta(12)$ can be directly identified with the quantity $3<\cos\Phi(r)>$ recently accumulated during numerical simulations [1,2] where $\Phi$ is the angle between two solvent dipole vectors and $\langle\cos\Phi(r)\rangle$ is defined as $\int g(r_{12},\Omega_{12})\cos\Phi_{12}d\Omega_{12}/\int d\Omega_{12}$. The law (1.1) has been established for pure dipolar solvents and spherical ions but is in fact valid in the general case, provided that the Debye screening length $\lambda_D=1/\kappa_D$ is much larger than any characteristic distance in the solvent. In particular, the hydrogen-bonding and tetrahedral structure of real water play a role only implicitly through the dielectric constant value $\varepsilon$. At this point, we draw the reader's attention to the fact that the corresponding $g_{ss}(12)$ correlation in *pure* solvent is dominated by a well-known $1/r^3$ asymptote in the D(12) component (see eq.5 below), so is overall comparatively much longer range than in dilute electrolyte. We will now illustrate and explain this behavior by playing with the HNC solution for the realistic SPC/E model of water mixed with different salts, bridging the gap between analytical limiting laws and brute-force numerical simulations involving one million atoms.

Ornstein-Zernike + HNC integral equation

The formalism and the numerical resolution of the OZ equation with approximate integral equations for molecular liquids and mixtures are well documented in the literature [8,9,10]. It is sufficient here to remind that the angular dependence of the different correlation functions like the pair distribution function $g_{ij}(\vec{r},\Omega)$ which couples pairs of particles belonging to species *i* and *j* is expressed as an expansion onto rotational invariants defined by 5 indices in the general case [13,14].

$$g_{ij}(\vec{r},\Omega) = \sum_{mnl\mu\nu} g_{ij}{}^{mnl}_{\mu\nu}(r)\Phi^{mnl}_{\mu\nu}(\hat{r},\Omega) \quad (1.2)$$

The expansion (1.2) is nothing but a generalization of (1.1) with $\Phi^{000}_{00}=1$, $\Phi^{110}_{00}\equiv-\sqrt{3}\Delta$, $\Phi^{112}_{00}\equiv\sqrt{3/10}D$,... (Blum's normalization) and may involve hundreds of projections for highly anisotropic angular couplings like those existing in water. Symmetries in molecular shape reduce the number of independent terms: the expansion contains a single term $g_{00}^{000}(r) \equiv g(r)$ for spherically-symmetric correlations like between atomic ions, $\mu\nu$=00 for linear particles like pure dipoles, $\mu$ and $\nu$ are even for $C_{2v}$ molecules like $H_2O$… The OZ equation which couples total $h \equiv g-1$ and direct $c$ correlation functions through spatial+angular convolution products is better expressed in the Fourier $q$-space and benefits from the expansion representation. First, each $f_{ij\ \mu\nu}^{\ mnl}(r)$ is Fourier-Hankel transformed to $\hat{f}_{ij\ \mu\nu}^{\ mnl}(q)$ and normalized by the cross density $(\rho_i\rho_j)^{1/2}$. Then, the OZ equation which becomes an algebraic product between $q$-projections is greatly simplified by privileging the intermolecular frame (that linked to the $\vec{q}$ vector) and introducing the new projection sets $\hat{f}_{ij\ \mu\nu;\chi}^{\ mn}(q)$ derived from the previous ones using the so-called Blum's $\chi$-transform, $\hat{f}_{ij\ \mu\nu;\chi}^{\ mn}(q) = \sum_l \begin{pmatrix} m & n & l \\ \chi & -\chi & 0 \end{pmatrix} \hat{f}_{ij\ \mu\nu}^{\ mnl}(q)$ [14]. Finally, for any mixture of anisotropic particles, the OZ equation reads [9]:

$$\hat{h}_{ij\ \mu\nu;\chi}^{\ mn}(q) = \hat{c}_{ij\ \mu\nu;\chi}^{\ mn}(q) + (-1)^\chi \sum_{k,p,\pi} (-1)^\pi \hat{h}_{ik\ \mu\pi;\chi}^{\ mp}(q) \hat{c}_{kj\ -\pi\nu;\chi}^{\ pn}(q) \quad (1.3)$$

where $\chi$ indices don't mix. Equation (1.3) is the key to extract and understand the low $q$ or long range $r$ behavior of the total correlations $h_{ij}$ from that of the direct ones $c_{ij}$, known as being $-v_{ij}/kT$ where $v_{ij}$ is the pair potential between species $i$ and $j$.

The exact OZ equation must be combined with a second, usually approximated relation coupling $h$'s and $c$'s in the $r$-space, a so-called integral equation. We use here the HNC closure relation which, despite its well documented underestimation of the H-bonding effect and dielectric constant $\varepsilon$ [15] [16], is sufficient in the present communication because it verifies the correct long-range behavior $c_{ij}=-v_{ij}/kT$ and accounts qualitatively for the different coupling effects in electrolytes. The technical details are described in the Supplementary Information (SI) in addition to ref.[11]. The following data correspond to NaCl electrolytes in the mM to M range.

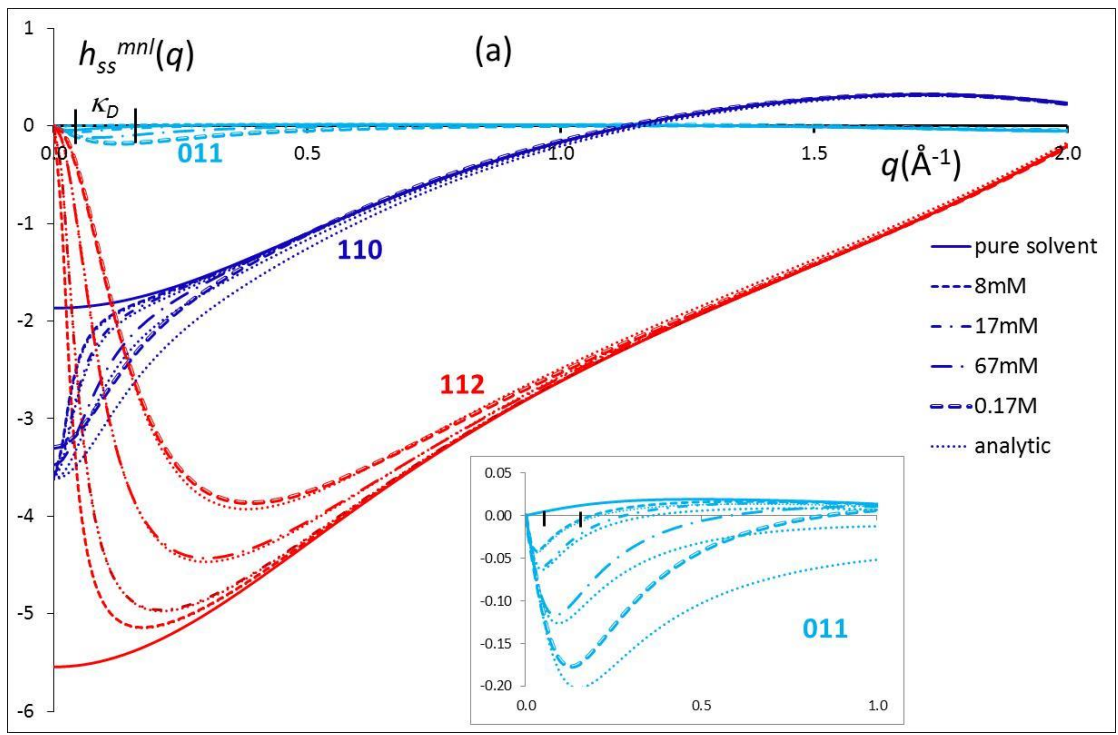

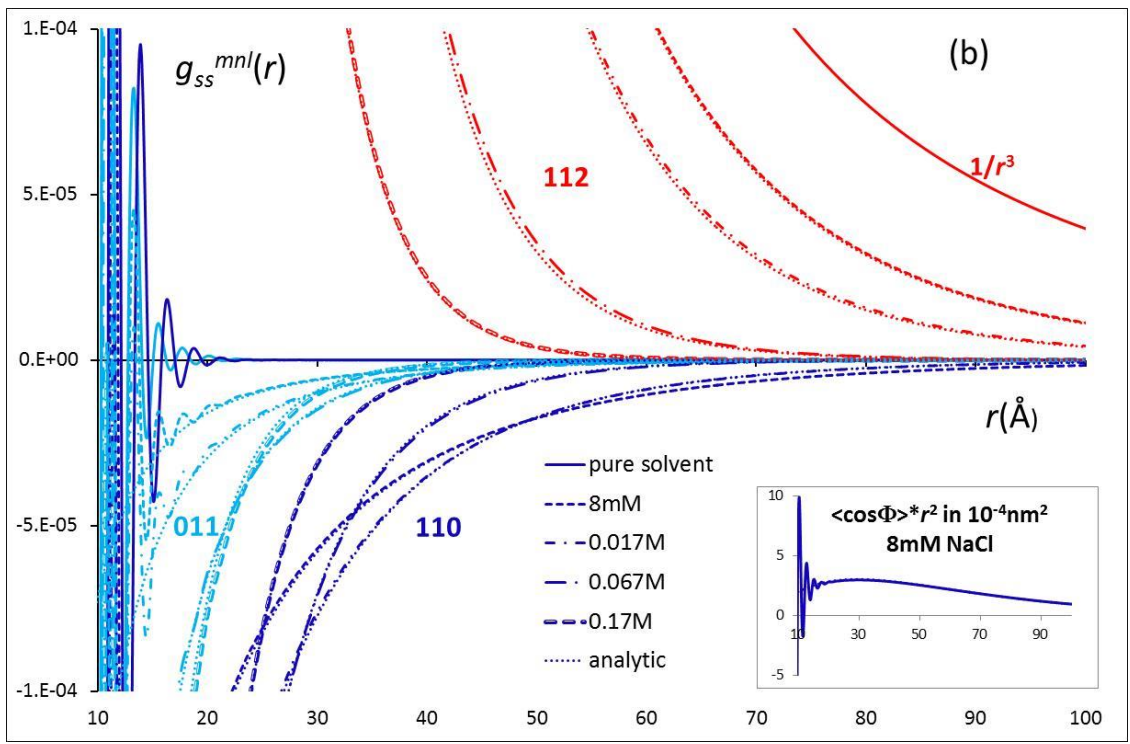

**Figure1.** First solvent-solvent projections $_{00}^{110}, _{00}^{112}, _{00}^{011}$ of the total correlation function $h$ in $q$-space (a) and in $r$-space (b) derived with the HNC integral equation for pure $H_2O$ and for NaCl aqueous electrolytes at different salinities. The value of the screening constant $\kappa_D$ for the two extreme salinities is marked on the $q$ horizontal axis. The insert in (a) magnifies the $_{00}^{011}$ projection (more precisely, its imaginary part). The insert in (b) plots $\langle \cos\Phi(r)\rangle * r^2 \equiv -\frac{1}{\sqrt{3}} h_{00}^{110}(r) * r^2$ for the 8mM salt, to be advantageously compared to figure 1 of ref. [2]. The dotted curves refer to the asymptotic laws (1.1), hardly distinguishable from the HNC curves at low salinity.

Figure 1a presents the first relevant projections of the solvent-solvent total correlation in the $q$-space. It clearly exhibits different behaviors at low $q$ between pure solvent and dilute, even very dilute, salt solutions. The salt curves deviate from the pure solvent ones for $q$ below $\kappa_D$, so at lower and lower $q$ values as the dilution increases. At zero $q$, all the $_{00}^{110}$ projections converge to an almost common value, independent of the salinity for the more dilute cases, while all $_{00}^{112}$ curves converge to exactly 0, in opposition to what happens in the pure solvent reference. These coupled behaviors simply illustrate the standard electrostatic screening due to presence of the ions and the well-known discontinuous behavior $\lim_{q\to 0}\lim_{\rho_{salt}\to 0} \neq \lim_{\rho_{salt}\to 0}\lim_{q\to 0}$ [17]. The corresponding curves in $r$ space are presented in figure 1b. The short-range $h_{ss}{}_{00}^{110}(r)$ and well-known $1/r^3$ tail of $h_{ss}{}_{00}^{112}(r)$ in pure solvent are replaced by screened-coulombic behaviors in presence of salt, in perfect agreement with the analytical laws (1.1). In order to be able to compare these results quantitatively with the recent simulation data of the literature, we have plotted in the insert of figure 1b and in figure 2 the quantity $<\cos\Phi(r)>\times r^2$ where $<\cos\Phi>\equiv -\frac{1}{\sqrt{3}} h_{ss}{}_{00}^{110}$, for the same salinities as in ref.2. The agreement is again spectacular (it happens coincidentally that the HNC $\varepsilon$ value of the SPC/E model, around 59, is very close to the true value of the TIP4P/2005 model used in ref.2). An apparent plateau in such plot results from the flat maximum of the function $re^{-\kappa_D r}$ (according to (1.1)), *not* from an hypothetical behavior of $\langle\cos\Phi(r)\rangle$ in $1/r^2$. A new projection $^{011}_{00}$, absent in the pure dipolar case and present in the more realistic SPC/E water model, displays similar qualitative salt-dependence, see figure 1. The same type of observation concerns the ion-solvent correlations as well, see SI.

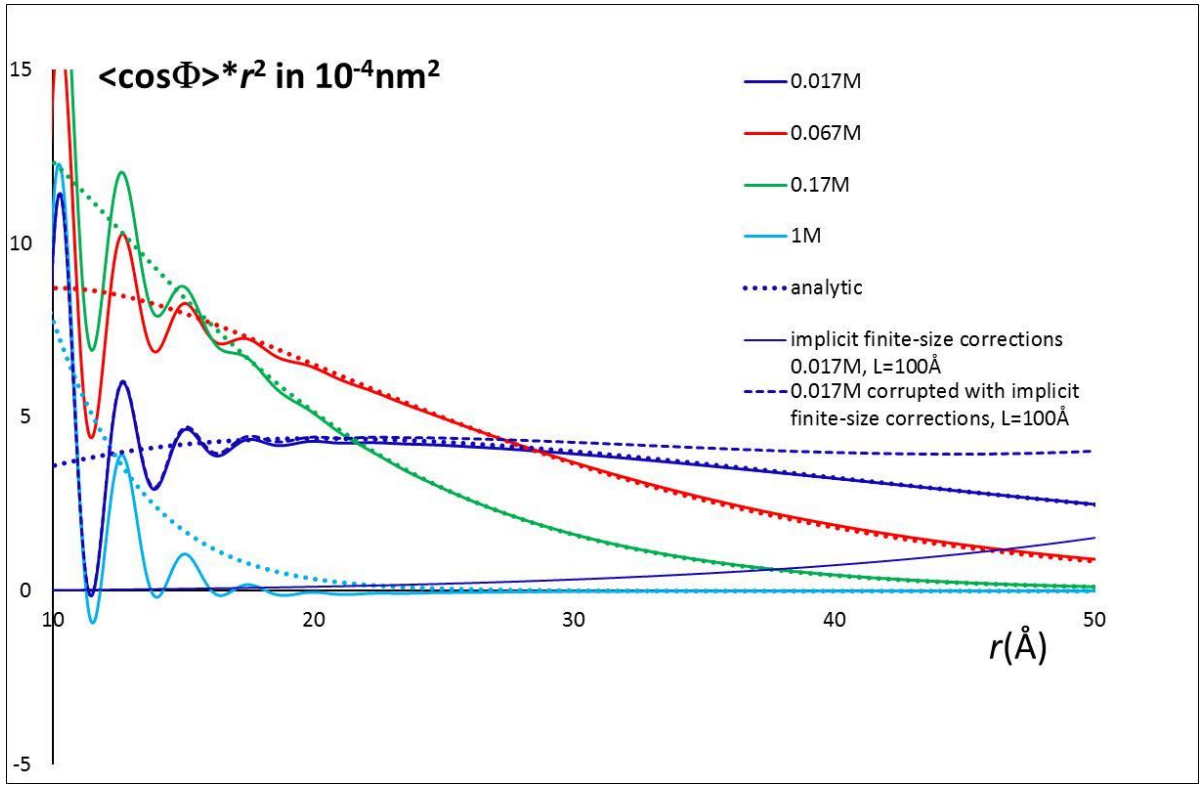

**Figure 2.** Solvent-solvent $^{110}_{00}$ projection of the total correlation function $h$ plotted as $\langle\cos\Phi(r)\rangle * r^2 \equiv -\frac{1}{\sqrt{3}} h^{110}_{00}(r) * r^2$ vs $r$ for different NaCl salinities. Solid lines: HNC; Dotted lines: asymptotic laws (1.1). This figure should be advantageously compared to the numerical simulation data in figure 4 of ref.[2]. For the 17mM NaCl case, the implicit corrections which are expected in a cubic simulation box of edge 10nm used in ref.2 are calculated as prescribed in ref.18.

Analytical laws

It is fruitful to understand the origin of the long-range asymptotes. It seems that the standard analysis has been somewhat ignored in the recent literature, so we take this opportunity to briefly recall its main steps. Everything starts from the OZ equation in $q$-space (1.3) which can be viewed as a matricial relation, for each $\chi$ value, between $\hat{h}$ and $\hat{c}$ projections [8][9]. Focusing on the low $q$ region, matrix inversion provides $\hat{h}$ from the known behavior $\hat{c} \approx -\hat{v}/kT$, independent of the density and salinity.

Let us first treat the pure solvent case. In the limit of zero $q$, the OZ equation couples $_{00}^{000}, _{00}^{110}, _{00}^{112}$ projections. The dipole-dipole $1/r^3$ contribution of the potential between H$_2$O molecules leads to the finite, non-zero value $\hat{c}_{ss}{}_{00}^{112}(0) = -\sqrt{30}y$. In pure solvent, and only in pure solvent, the Kirkwood relation expresses the dielectric constant $\varepsilon$ in terms of the $\hat{h}_{ss}{}_{00}^{110}(0)$ projection through:

$$\frac{(\varepsilon-1)(2\varepsilon+1)}{3\varepsilon} = 3y\left(1 - \frac{1}{\sqrt{3}}\hat{h}_{\text{pure}\,ss}{}_{00}^{110}(0)\right) \qquad (1.4)$$

These two relations combined with the two OZ equations (one for $\chi=0$, one for $\chi=1$) lead to:

$$\hat{h}_{\text{pure}\,ss}{}_{00;0}^{11}(0) = \frac{\varepsilon-1}{3y\varepsilon} - 1 \; ; \; \hat{h}_{\text{pure}\,ss}{}_{00;1}^{11}(0) = 1 - \frac{\varepsilon-1}{3y}$$

$$\hat{h}_{\text{pure}\,ss}{}_{00}^{112}(0) = -\sqrt{\tfrac{10}{3}}\frac{(\varepsilon-1)^2}{3y\varepsilon} \; ; \; h_{\text{pure}\,ss}{}_{00}^{112}(r) \underset{r\to\infty}{\approx} \sqrt{\tfrac{10}{3}}\frac{(\varepsilon-1)^2}{4\pi\rho_s y\varepsilon r^3} \qquad (1.5)$$

This already mentioned $r^{-3}$ asymptote of $h_{\text{pure}\,ss}{}_{00}^{112}(r)$, present in pure solvent only, perfectly fits the tail of the HNC curve in figure 1b.

How is the OZ equation modified in presence of *dilute* salt? As the salt concentration goes to small values, ions play a role only through the long range $1/r$ ion-ion and $1/r^2$ ion-dipole potentials. That means that the only relevant ionic contributions to the direct correlation functions are the following divergences at low $q$:

$$\hat{c}_{ij\ 00}^{\ 000}(q) = -\frac{q_i q_j}{q^2}$$

$$\hat{c}_{is\ 00}^{\ 011}(q) = -\hat{c}_{si\ 00}^{\ 101}(q) = i\frac{q_i\sqrt{3y}}{q} \qquad (1.6)$$

where $q_i = \sqrt{4\pi L_B \rho_i} Z_i$ is proportional to the valency $Z_i$ of ion $i$. The screening constant in *vacuum* reads simply $\kappa = \sqrt{\sum_i q_i^2}$ and is related to that in solvent by $\kappa_D^2 = \kappa^2/\varepsilon$. Since in that limit the ions are considered as (point-like) spherically-symmetric particles, only the $\chi=0$ OZ relation is perturbed by the presence of ions. OZ matrix inversion leads to the Debye-Hückel ion-ion correlations $\hat{h}_{ij}(q) = -\frac{q_i q_j/\varepsilon}{q^2 + \kappa_D^2}$, as expected, and to the new solvent-solvent correlations (see technical details in SI):

$$\hat{h}_{ss\ 00;0}^{\ 11}(q) = \frac{\varepsilon-1}{3y\varepsilon}\frac{q^2+\kappa^2}{q^2+\kappa_D^2} - 1 = \hat{h}_{\text{pure}\,ss\ 00;0}^{\ 11}(0) + \frac{(\varepsilon-1)^2}{3y\varepsilon}\frac{\kappa_D^2}{q^2+\kappa_D^2}$$

$$\hat{h}_{ss\ 00}^{\ 110}(q) = \hat{h}_{\text{pure}\,ss\ 00}^{\ 110}(0) - \frac{(\varepsilon-1)^2}{3\sqrt{3}y\varepsilon}\frac{\kappa_D^2}{q^2+\kappa_D^2} \qquad (1.7)$$

$$\hat{h}_{ss\ 00}^{\ 112}(q) = \hat{h}_{\text{pure}\,ss\ 00}^{\ 112}(0) + \sqrt{10/3}\frac{(\varepsilon-1)^2}{3y\varepsilon}\frac{\kappa_D^2}{q^2+\kappa_D^2}$$

The new, ion-induced lorentzian contributions, independent of the salinity at zero $q$, perfectly fit the HNC curves in figure 1a, except for the highest salinity. One notes in particular that the new $_{00}^{112}$ term exactly balances the pure solvent value (1.5) at $q=0$. This automatically implies the disappearance, due to ionic screening, of the corresponding $1/r^3$ in $r$ space and the replacement by screened coulombic behaviors, both for $_{00}^{110}$ and $_{00}^{112}$ projections, in perfect agreement with the original limiting laws (1.1). Similar laws can be found for the ion-solvent correlations as well ([3] and SI).

In conclusion, the present work, mixing integral equation theory and analytical treatment, explains the recent large scale numerical simulation data in terms of well-established screened coulombic generic effects. Focusing solely on the <cosΦ> projection is somewhat misleading: as regards to the solvent orientational correlations, the main effect of ions consists in killing the long range $1/r^3$ [112] dipolar contribution, and thus in decreasing the solvent-solvent correlation range in that symmetry which indeed dominates the long-range interactions.

We would like to add a few remarks:

1) Going through the $q$ space is the natural and practical way to resolve the OZ equation but may prevent one from understanding the physical picture behind the mathematics. It is possible to stay in the $r$ space and solve pertubatively the OZ equation to get the correction to the solvent-solvent correlation $<\cos\Phi(r)>$ to first order in the salt concentration, namely, summing over ion types:

$$\sum_i \rho_i \int d\vec{r}_i c_{si}(\vec{r}_{1i},\Omega_1) c_{is}(\vec{r}_{i2},\Omega_2) = L_B^2 \mu_s^2 \sum_i \rho_i Z_i^2 \int d\vec{r}_i \left(\frac{\hat{\mu}_1.\hat{r}_{1i}}{r_{1i}^2}\right)\left(\frac{\hat{\mu}_2.\hat{r}_{2i}}{r_{2i}^2}\right) \qquad (1.8)$$

Thus, this extra correlation comes from two solvent molecules 1 and 2 feeling the Coulombic interaction with the *same ion*. Performing analytically the 3D convolution leads to the $1/r$ solvent-solvent correlation $\frac{\kappa^2 L_B \mu_s^2}{2r_{12}}(\hat{\mu}_1.\hat{\mu}_2 - (\hat{r}_{12}.\hat{\mu}_1)(\hat{r}_{12}.\hat{\mu}_2))$ which recovers the limiting law (1.1) expanded to first order in $\rho_{salt}$ (second order in $\kappa_D$) with $\varepsilon$-1 replaced by its ideal quantity $3y<<1$. The correct factor in $(\varepsilon-1)^2/\varepsilon$ appearing in (1.1) means that solvent 1 should be first convoluted to one, two… intermediate solvent neighbors through the pure solvent correlations before reaching the ion (same for solvent 2) while the screening factor $\exp(-\kappa_D r)$ results from Debye-Hückel ion-ion correlations through standard chain diagrams [3]. Along a similar route, we note a recent attempt to derive the ion-induced solvent correlations using a dipolar gas model for the solvent [19]. This approach does yield the correct $\exp(-\kappa r)/r$ behavior for $<\cos\Phi(r)>$, but with a wrong prefactor suffering from the same ideal gas approximation of the dielectric constant, $\varepsilon-1=3y$; this approximation was somehow compensated in the paper by the ad-hoc addition of an "Onsager local field factor".

2) While the Kirkwood formula (1.4) is valid for ion-free solvents, the corresponding formula for electrolytes reads $\varepsilon-1=3y\left(1-\frac{1}{\sqrt{3}}\hat{h}_{ss\ 00}^{110}(0)\right)$ [6]. Since $\varepsilon$ is a continuous function of the salinity, comparison of both expressions automatically implies a discontinuous step of $\hat{h}_{ss\ 00}^{110}(0)$ going from pure solvent to infinitely dilute electrolyte, illustrated in figure 1a.

3) From the previous remark, one would conclude that if an experiment really measures the Fourier transform of $<\cos\Phi(r)>$ at exactly zero $q$ (integral of $<\cos\Phi(r)>*r^2$), the result would present a discontinuous step from pure solvent to electrolyte. If, more realistically, the experiment measures the Fourier transform at low but finite $q$, as in light scattering techniques, one would get a continuous behavior in $\frac{\kappa_D^2}{q^2+\kappa_D^2}$. At 90° scattering angle, the 50% onset at $\kappa_D \approx q$ would fall in the 50μM salinity range. This could be advantageously compared to the recent femtosecond elastic second harmonic scattering data [1].

4) An attentive reader may have detected a small but clear disagreement between figure 2 and the corresponding original figure 4 in reference [2] about the 17mM NaCl case. This illustrates the presence of *implicit* finite-size corrections in the numerical simulation data due to the environment around the images in the neighboring cells within the periodic boundary conditions. That happens when the cubic cell size $L$ is not large enough compared to the characteristic correlation length (here, $L$=10nm while $\lambda_D$=2nm). Recently, it has been shown how to evaluate such corrections with great precision in dipolar solvents [18]. Here, the simple asymptotic law makes the calculation analytical. If $<\cos\Phi(r)>$ behaves as $A*\exp(-\kappa_D r)/r$, the influence of the first 6 neighbors adds the extra term $6A\dfrac{\sinh \kappa_D r}{\kappa_D rL}e^{-\kappa_D L}$, see SI. When this expression is added to the HNC/limiting law curve in order to emulate raw simulation data corrupted with such finite-size corrections, one recovers, this time very nicely, the curve of ref. [2], see figure 2. The apparent $1/r^2$ behavior (plateau in fig.2) that was observed is therefore an artefact.

Supporting Information: HNC integral equation and limiting law computing details; ion-ion and ion-solvent correlations (PDF file)

# Supplementary materials: Screened coulombic orientational correlations in dilute aqueous electrolytes

Luc Belloni[1*], Daniel Borgis[2,3] and Maximilien Levesque[3]

[1] *LIONS, NIMBE, CEA, CNRS, Université Paris-Saclay, 91191 Gif-sur-Yvette, France*

[2] *Maison de la Simulation, USR 3441 CNRS-CEA-Université Paris-Saclay, 91191 Gif-sur-Yvette, France*

[3] *PASTEUR, Département de chimie, École normale supérieure, PSL University, Sorbonne Université, CNRS, 75005 Paris, France*

## Expansion in rotational invariants

Using Blum's notation and normalization [1,2], the expansion (eq.2 of the main text) reads:

$$g_{ij}(\vec{r},\Omega) = \sum_{mnl\mu\nu} g_{ij\ \mu\nu}^{\ mnl}(r)\Phi_{\mu\nu}^{mnl}(\hat{r},\Omega) \qquad (0.1)$$

where the rotational invariants, independent of the choice for the reference frame, are:

$$\Phi_{\mu\nu}^{mnl}(\Omega_1,\Omega_2,\hat{r}_{12}) = f_m f_n \sum_{\mu',\nu',\lambda'} \begin{pmatrix} m & n & l \\ \mu' & \nu' & \lambda' \end{pmatrix} R_{\mu'\mu}^m(\Omega_1) R_{\nu'\nu}^n(\Omega_2) R_{\lambda'0}^l(\hat{r}_{12}) \qquad (0.2)$$

$\begin{pmatrix} m & n & l \\ \mu' & \nu' & \lambda' \end{pmatrix}$ is the usual 3-j-symbol; $f_m = (2m+1)^{1/2}$. The $R_{\mu'\mu}^m(\Omega)$ are Wigner generalized spherical harmonics (definition and notation from Messiah [3]). Comparison with the historical notation as in eq.1 of the main text:

$$g(\vec{r},\Omega) = g_S(r) + g_\Delta(r)\Delta(\Omega) + g_D(r)D(\Omega) + \ldots \qquad (0.3)$$

means that $g_S \equiv g_{00}^{000}$; $g_\Delta \equiv -\sqrt{3} g_{00}^{110} \equiv 3<\cos\Phi>$; $g_D \equiv \sqrt{3/10}\, g_{00}^{112}$.

## HNC integral equation

The Ornstein-Zernike equation has been solved with the HNC closure for mixtures of spherical $Na^+$, $Cl^-$ ions and SPC/E water molecules. The Lennard-Jones parameters for the ions are $\sigma_+ = 2.583$Å, $\sigma_- = 4.401$Å, $\varepsilon_+ = \varepsilon_- = 0.4165$kJ/mol but we repeat that in the present context of limiting laws valid at low salinity, the precise values are irrelevant. The basis of invariants has been truncated at $m,n \leq n_{max}$ with $n_{max}=0$ for the ions and $=4$ for $H_2O$, corresponding to 1, 9 and 250 independent projections for each ion-ion, each ion-water and water-water sets, respectively. The numerical resolution follows the powerful technique described in [4]. The HNC dielectric constant of this pure SPC/E water model is 59 (while the exact $\varepsilon$ is around 72). It is fortuitous that this value is close to that of the TPIP4P/2005 model used in the numerical simulations [5]. Since the limiting laws depend on the ionic strength and $\varepsilon$ only, that allows for direct, quantitative comparison.

In a few cases, we have added a small dipole to the ions in order to illustrate the general character of the limiting laws. With $n_{max}=1$ for the ions, the new number of independent projections becomes 4 for ++ and --, 5 for +- and 30 for each ion-$H_2O$.

## Ornstein-Zernike inversion and limiting laws

The OZ equation couples total $h$ and $c$ direct correlation functions. The idea of the present analytical treatment consists to extract the long distance $r$ or low $q$ behavior of $h$ from that of $c$ or of the pair potential $v$ since $c \approx -v/kT$ in that regime. The OZ convolution product becomes in Fourier space an algebraic product between the Blum's $\chi$-projections, see eq.3 in the main text. In practice, that consists, for each value of the indices $\chi$, to inverse a matrix of the form

$$S_\chi = 1 + (-1)^\chi \hat{h}_\chi = \left(1 - (-1)^\chi \hat{c}_\chi\right)^{-1} \ [6].$$

Pure solvent:

For pure solvent, at exactly zero $q$, the projections like $mn=01$ vanish as $iq$ and the projections $mn=00$ and 11 are coupled via 2*2 and 1*1 matrices for $\chi=0$ and 1, respectively ($q=0$ implicit):

$$S^{pure}_{\ 0} = \begin{pmatrix} 1 + \hat{h}_{pure\,ss}^{\ 00}{}_{00;0} & 0 \\ 0 & 1 + \hat{h}_{pure\,ss}^{\ 11}{}_{00;0} \end{pmatrix} = \begin{pmatrix} 1 - \hat{c}^{00}_{00;0} & 0 \\ 0 & 1 - \hat{c}^{11}_{00;0} \end{pmatrix}^{-1}$$

$$S^{pure11}_{\ \ 00;1} = 1 - \hat{h}_{pure\,ss}^{\ 11}{}_{00;1} = \left(1 + \hat{c}^{11}_{00;1}\right)^{-1} \tag{0.4}$$

The dipole-dipole contribution to the pair potential implies $\hat{c}_{ss}^{\ 112}{}_{00} = -\sqrt{30}\,y$ and the $\chi$-transforms $\hat{c}_{ss}^{\ 11}{}_{00;0} = -\frac{1}{\sqrt{3}}\hat{c}_{ss}^{\ 110}{}_{00} - 2y$ and $\hat{c}_{ss}^{\ 11}{}_{00;1} = \frac{1}{\sqrt{3}}\hat{c}_{ss}^{\ 110}{}_{00} - y$. Using the Kirkwood equation (eq.4 of the main text) which relates dielectric constant $\varepsilon$ and $\hat{h}_{ss}^{\ 110}{}_{00}(0)$ projection leads to the final result (eq.5 of the main text):

$$\hat{c}_{ss}^{\ 11}{}_{00;0} = 1 - \frac{3y\varepsilon}{\varepsilon - 1}; \quad S^{pure11}_{\ \ 00;0} = \frac{\varepsilon - 1}{3y\varepsilon}; \quad S^{pure11}_{\ \ 00;1} = \frac{\varepsilon - 1}{3y} \tag{0.5}$$

Dilute electrolyte:

The presence of ions induces at low $q$ diverging terms in $1/q^2$ for ion-ion and $i/q$ for ion-solvent direct correlations (eq.6 of the main text), which concern the $\chi=0$ OZ equation. In the limit of low salt concentration and small $q$, these are the only adding terms which affect the OZ equation. For a binary salt, each line and column of $S_{\chi=0}$ contains successively $\substack{m \\ i} = \substack{0 \\ +}, \substack{0 \\ -}, \substack{0 \\ s}, \substack{1 \\ s}$. The OZ matrix becomes (the indices $\chi=0$ and $\mu\nu=00$ are dropped for clarity):

$$S = \begin{pmatrix} 1+\hat{h}^{00}_{++} & \hat{h}^{00}_{+-} & 0 & \hat{h}^{01}_{+s} \\ \hat{h}^{00}_{-+} & 1+\hat{h}^{00}_{--} & 0 & \hat{h}^{01}_{-s} \\ 0 & 0 & 1+\hat{h}^{00}_{ss} & 0 \\ \hat{h}^{10}_{s+} & \hat{h}^{10}_{s-} & 0 & 1+\hat{h}^{11}_{ss} \end{pmatrix} = \begin{pmatrix} 1+\dfrac{q_+q_+}{q^2} & \dfrac{q_+q_-}{q^2} & 0 & i\dfrac{q_+\sqrt{3y}}{q} \\ \dfrac{q_+q_-}{q^2} & 1+\dfrac{q_-q_-}{q^2} & 0 & i\dfrac{q_-\sqrt{3y}}{q} \\ 0 & 0 & 1-\hat{c}^{00}_{ss} & 0 \\ -i\dfrac{q_+\sqrt{3y}}{q} & -i\dfrac{q_-\sqrt{3y}}{q} & 0 & \dfrac{3y\varepsilon}{\varepsilon-1} \end{pmatrix}^{-1} \quad (0.6)$$

As usual in charged systems, the effect of the divergences is neutralized by ... the electroneutrality condition $\rho_+ Z_+ + \rho_- Z_- = 0$. The matrix inversion is straightforward. For ion-ion, one gets:

$$\hat{h}^{00}_{ij}(q) = -\frac{q_i q_j / \varepsilon}{q^2 + \kappa_D^2}$$

$$h_{ij}(r) = -\frac{Z_i Z_j L_B}{\varepsilon r} e^{-\kappa_D r} \quad (0.7)$$

which is nothing but the standard Debye-Hückel law. In Fourier space, the correlation disappears as soon as $q \gg \kappa_D$ while it reaches a constant, independent of the salinity, at $q=0$. For a symmetrical electrolyte like NaCl, $S_{++}=S_{+-}=S_{--}=1/2$! That just means that the osmotic pressure follows the perfect gas law, $\Pi = (\rho_+ + \rho_-)kT = 2\rho_{salt} kT$ [7].

As for the new solvent-solvent total correlations, eq.(0.6) leads in $q$ space to eq.7 of the main text. An inverse Fourier-Hankel transform yields the Hoye and Stell's limiting law (eq.1 of the main text), in particular:

$$\langle \cos \Phi(r) \rangle = \frac{(\varepsilon-1)^2 \kappa_D^2}{36\pi \rho_s y \varepsilon} \frac{e^{-\kappa_D r}}{r} \quad (0.8)$$

Finally, the orientation of the water molecules around the ions is described by the $h^{011}_{is\ 00}$ projection:

$$\hat{h}^{01}_{is\ 00;0}(q) \equiv -\frac{1}{\sqrt{3}} \hat{h}^{011}_{is\ 00}(q) = iq \frac{\varepsilon-1}{\sqrt{3y\varepsilon}} \frac{q_i}{q^2 + \kappa_D^2}$$

$$h^{011}_{is\ 00}(r) = -Z_i \mu_s L_B \frac{\varepsilon-1}{3y\varepsilon} \frac{1+\kappa_D r}{r^2} e^{-\kappa_D r} \quad (0.9)$$

The limiting law in $r$ space has been given by Hoye and Stell (note that $\Phi^{011}_{00} = -\hat{\mu}_s . \hat{r}_{i \to s}$). The agreement with the HNC numerical data is illuminating, see figure S1. Note that the slope at the origin in $q$-space diverges at infinite dilution: in the case of one single ion immersed in bulk water, the ion-solvent orientational total correlations become unscreened and behave as $1/q$ or $1/r^2$.

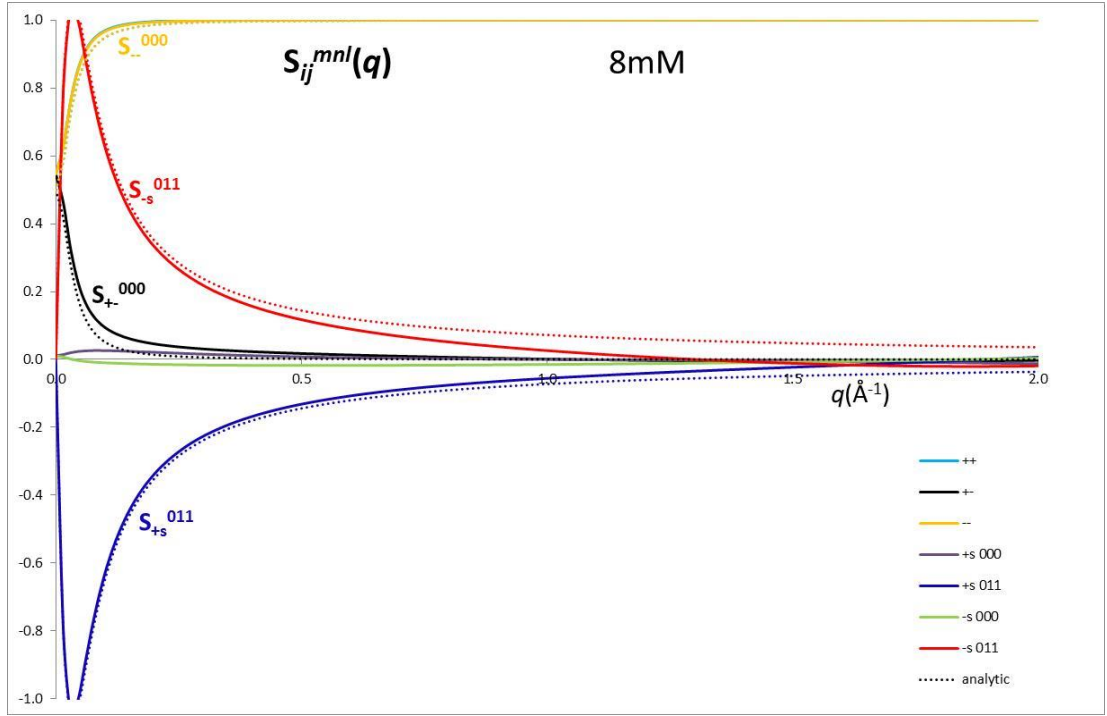

*Figure S1: HNC curves (solid lines) and asymptotic laws (dotted lines) for the ion-ion and ion-solvent total correlation functions. The 011 curves must be understood as the imaginary part of the corresponding functions.*

The SPC/E site-site model for the water molecules used in the present HNC theory goes beyond the simple dipolar picture. In particular, it involves the first extra projection $h_{ss\,00}^{011}$ for the solvent-solvent correlations. Since this one vanishes as $iq$ at low $q$, the direct correlation matrix in (0.6) must be enriched with so far neglected $\hat{c}_{is}^{\,00}$ and $\hat{c}_{ss}^{\,01}(q) \approx iq\hat{c}_{ss}^{\,01}{}'$ contributions. The final expression for symmetrical salt reads:

$$S_{ss}^{\,01}(q) \approx S_{\text{pure}\,ss}^{\,01}(q) + iq\,\frac{\varepsilon-1}{(1-\hat{c}_{ss}^{\,00})3y\varepsilon}\,\frac{(\varepsilon-1)\hat{c}_{ss}^{\,01}{}'\kappa_D^{\,2} + \sqrt{3/2}\,y\varepsilon(\hat{c}_{+s}^{\,00} - \hat{c}_{-s}^{\,00})\kappa_D}{q^2 + \kappa_D^{\,2}} \quad (0.10)$$

As seen in the insert of fig.1 of the main text, this original limiting law reproduces very nicely the numerical HNC curves for the different low salinities.

<u>Non-symmetrical ions:</u>

We have added here small dipoles to the ions, $\mu_+$=0.2Å and $\mu_-$=0.15Å, in order to illustrate the non-sensitivity of the long-range limiting laws on the details of the ions. The HNC solution is given in figure S2 for 8mM NaCl. New ion-solvent projections like $^{011}$ or $^{110}$ now appear, which were completely absent in the case of spherical ions. They illustrate the change in the orientation of the water molecules relative to the ions inside the solvation layers. Despite this clear effect, the solvent-solvent orientational order $<\cos\Phi(r)>$ is almost identical to the reference case, see the insert.

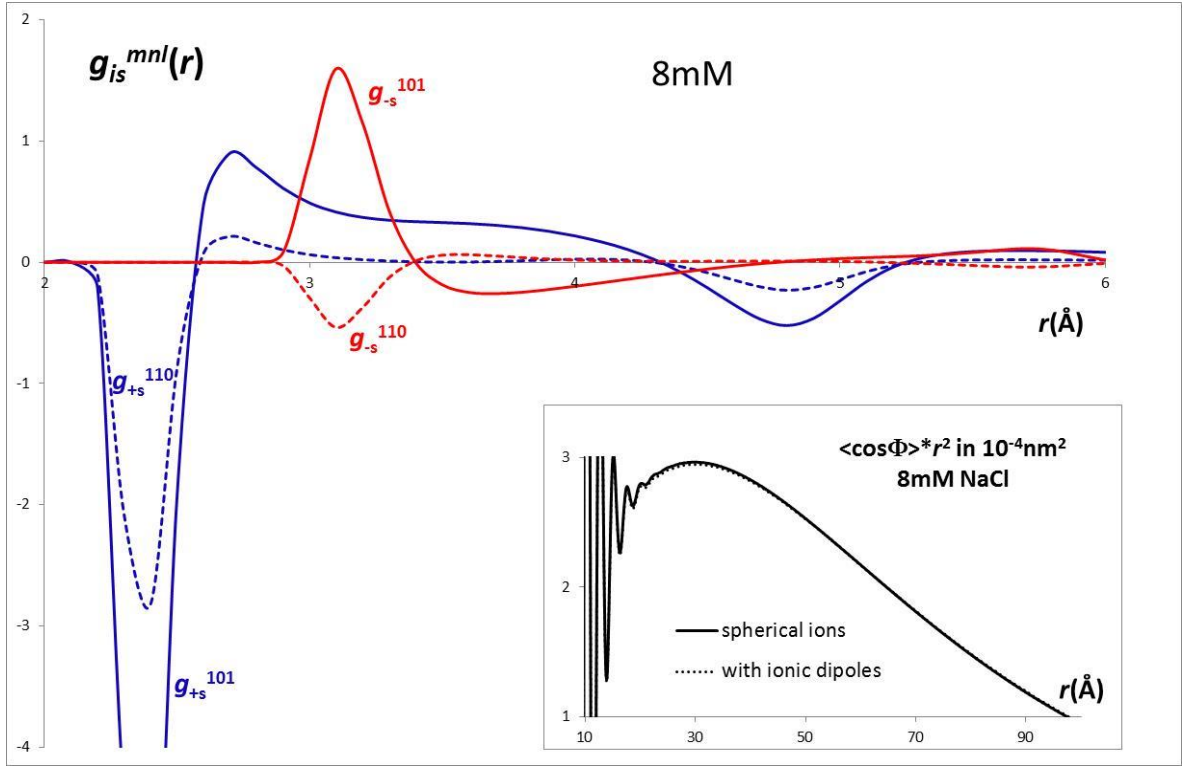

*Figure S2: HNC solution for 8mM NaCl in presence of ionic dipoles, $\mu_+=0.2$Å and $\mu_-=0.15$Å. Main figure: New ion-solvent projections $^{101}$ and $^{110}$ illustrate the change in the orientation of the water molecules inside the solvation layer. Insert: the long distance solvent-solvent orientational order is unchanged.*

### Implicit finite-size corrections in numerical simulation

Within the periodic boundary conditions, the pair distribution function measured inside the central cell of edge $L$ is perturbed by the environment around the neighboring images. The systematic and detailed treatment of such corrections for molecular systems has been described recently [8]. Here, we are interested in the solvent-solvent $^{110}$ or $<\cos\Phi>$ projection in figure 2 of the main text. Within the superposition approximation, the first 6 neighbors add an extra, undesired correction [8]:

$$\Delta h_{ss00}^{110}(r) = 6\int_0^\pi h_{ss00}^{110}\left(\sqrt{r^2+L^2-2rL\cos\theta}\right) \tfrac{1}{2} d\cos\theta \qquad (0.11)$$

When the true correlation takes a screened coulombic Yukawa form $Ae^{-\kappa_D r}/r$, the integration can be performed analytically:

$$\Delta h_{ss00}^{110}(r) = 6A\frac{\sinh \kappa_D r}{\kappa_D rL}e^{-\kappa_D L} \qquad (0.12)$$

When this correction is added to the bare limiting law for the case $L$=10nm, $\rho_{salt}$=17mM, the agreement with the numerical simulation data becomes spectacular (figure 2 in the main text and figure 4 of [5]).